# HYBRID MESSAGE-EMBEDDED CIPHER USING LOGISTIC MAP

Mina Mishra[1] and Dr. V.H. Mankar[2]

[1]Ph. D. Scholar, Department of Electronics & Telecommunication, Nagpur University, Nagpur, Maharashtra, India
minamishraetc@gmail.com
[2]Senior Faculty, Department of Electronics Engineering, Government Polytechnic, Nagpur, Maharashtra, India
vhmankar@gmail.com

## ABSTRACT

*The proposed hybrid message embedded scheme consists of hill cipher combined with message embedded chaotic scheme. Message-embedded scheme using non-linear feedback shift register as non-linear function and 1-D logistic map as chaotic map is modified, analyzed and tested for avalanche property and strength against known plaintext attack and brute-force attack. Parameter of logistic map acts as a secret key. As we know that the minimum key space to resist brute-force attack is 2100, and it is observed from analysis that key space of the discussed method is lesser than 2100. But the identifiability test concludes that the scheme consists of identifiable keys which are sufficient condition to resist brute-force attack for chaotic ciphers.*

*A complete file can be encrypted and decrypted successfully by the method that assures security against brute force attack. It is also concluded that the scheme has an average key sensitivity.*

## 1. INTRODUCTION

Chaos theory [1] has been received in the last two decades a great deal of attention from the scientific community. Remarkable research efforts have been invested in recent years, trying to export concepts from physics and mathematics into real world engineering applications. Among the most promising applications of chaotic systems [2] is their use in the field of chaotic encryption where the utilization of nonlinearities and the forcing of the dynamical system to a chaotic state will fulfill the basic cryptographic requirements. Due to the nonlinear mechanisms [3] that lead to a chaotic behavior, this one is too difficult to predict by analytical methods without the secret key (initial conditions and/or parameters) being known. This would reduce a potential attack to one category that of a brute force attack, in which any attempt to crack the key depends directly upon how long the key is. Chaos is a particular state of a nonlinear dynamical system and appears only in certain conditions, e.g. for certain values of the system parameters and only in dynamical systems characterized by continuous values.

Chaos-based encryption scheme [4] [5] is an efficient and hot research aspect considering virtues of chaotic sequence. It is well known that chaotic sequences are pseudorandom, non-periodic, unpredictability, and especially sensitive to initial parameters, or say good avalanche effect. These features are very important for modern cipher algorithms, especially for stream ciphers.

One of the most promising chaotic cryptosystem scheme known as hybrid message- embedding [6] is explained and illustrated with the help of fig 1.

**(i) Transmitter and Encryption:** Plain text is encrypted using traditional method (hill cipher) and the encrypted plaintext is again encrypted by an encryption rule which uses non-linear function and the state generated by the chaotic system in the transmitter. The scrambled output is inputted further to the chaotic system [7] such that the chaotic dynamics [8] is changed continuously in a very complex way. Then another state variable of the chaotic system in the transmitter is transmitted through the channel.

**(ii) Receiver and Decryption:** Recovery of the plaintext is done by decrypting the input (cipher text) using reverse process of encryption, as used in the transmitter.

In this scheme state X is not directly transmitted through the noise interference channel but quantity Y is available at the output of transmitter.

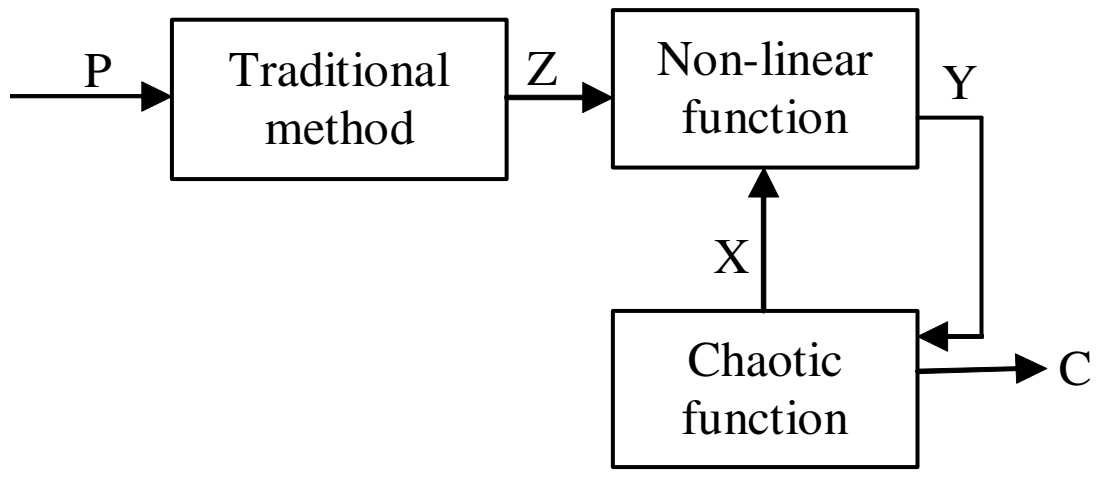

(a) Transmitter (encryptor)

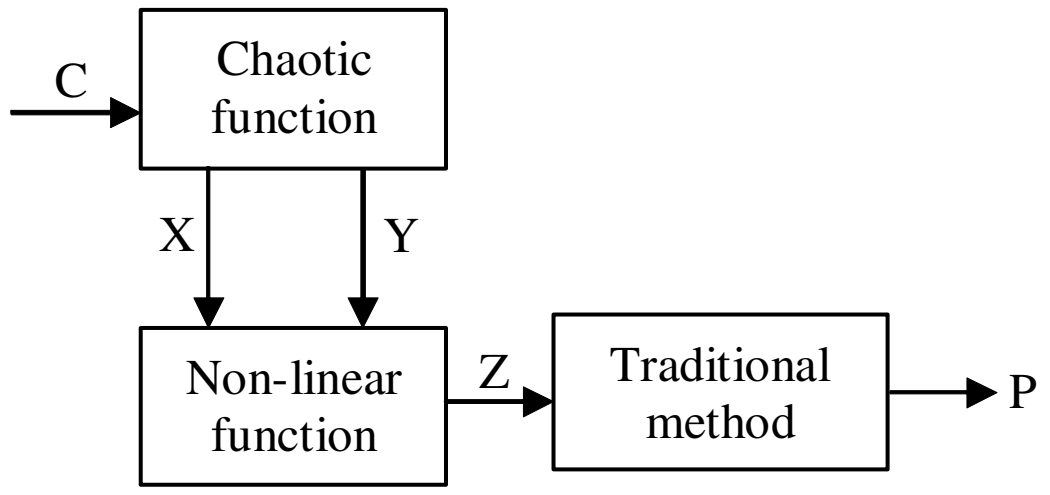

(b) Receiver (decryptor)

Figure 1. Hybrid message-embedded chaotic cryptosystem

P: plaintext; C: cipher text; X: state of chaotic function;
Z &Y: Intermediate encrypted plaintext;

An essential issue for the validation of any cryptosystems is the cryptanalysis [9] that is the study of attacks against cryptographic schemes in order to reveal their possible weakness. A fundamental assumption in cryptanalysis, first stated by A. Kerkhoff in 1883, is that the adversary knows all the details of the cryptosystem, including the algorithm and its implementation, except the secret key, on which the security of the cryptosystem must be entirely based.

The various cryptanalytic attacks are:

**(i) Cipher text-only attack:** The attacker possesses a string of cipher text.

**(ii) Known plain text:** The attacker possesses a string of plain text, P, and the corresponding cipher text, C.

**(iii) Chosen plain text:** The attacker has obtained temporary access to the encryption machinery. Hence he/she can choose a plain text string, P, and construct the corresponding cipher text string, C.

**(iv) Chosen cipher text:** The attacker has obtained temporary access to the decryption machinery. Hence he/she can choose a cipher text string, C, and construct the corresponding plain text string, P.

**(v) Brute Force Attack:** A brute force attack is the method of breaking a cipher by trying every possible key. The brute force attack is the most expensive one, owing to the exhaustive search.
There are some other specialized attacks, like, differential and linear attacks. Differential cryptanalysis is a kind of chosen-plaintext attack aimed at finding the secret key in a cipher. It analyzes the effect of particular differences in chosen plaintext pairs on the differences of the resultant cipher text pairs. These differences can be used to assign probabilities to the possible keys and to locate the most probable key. Linear cryptanalysis [10] is a type of known-plaintext attack, whose purpose is to construct a linear approximate expression of the cipher under study. It is a method of finding a linear approximation expression or linear path between plaintext and cipher text bits and then extends it to the entire algorithm and finally reaches a linear approximate expression without intermediate value.

From the crypto graphical point of view [11], the size of the key space should not be smaller than 2100 to provide a high level security so that it can resist all kind of Brute force attack. To get a number of keys with 2100 (approx. 1030), in chaotic system the resolution must be 10-15, but, it may be possible that thousands of keys would become equivalent with that resolution, unless, there is a strong sensitivity to parameter mismatch. The quicker the brute force attack, the weaker the cipher. Whether brute force attacks gets succeeded or not depends on the key space size of the cipher and on the amount of computational power available to the attacker. However, this requirement might be very difficult to meet by proposed cipher because the key space does not allow for such a big number of different strong keys. The brute force attack is the most expensive one, owing to the exhaustive search. A fundamental issue of all kinds of cryptosystem is the key. No matter how strong and how well designed the encryption algorithm might be, if the key is poorly chosen or the key space is too small, the cryptosystem will be easily broken. Unfortunately, proposed chaotic cryptosystem has a small key space region and it is non-linear because all the keys are not equally strong.

A cryptanalytic method, known as output equality based on the identifiability concept cited in [12], is the solution to the problem of less key space in chaotic ciphers. It is found that in chaotic ciphers, there exists a unique solution for a particular input for certain domain of values of parameters. The response of any system to a particular input is the solution of that particular system and it contains all the information about the parameters of system. This type of analysis is also known as parametric analysis. Identifiability concept fulfills the necessary condition but not sufficient as the developed cryptosystems must be tested for sensitivity and other statistical tests to result in a robust cipher.

This work aims to present method for encryption using Logistic chaotic map that provides security against Brute-Force attack and improved plaintext sensitivity and key sensitivity. The analysis result concludes that the proposed encryption algorithm shows improvement in plaintext sensitivity and key sensitivity property. Method is found to resist known plaintext attack for almost all the selected keys as shown in the analysis table for available first five characters of plaintext string. All the available string of plaintext must be the starting characters of plaintext then only there exists some probability to find out the key by known plaintext attack. Conclusion about the identifiability of almost all chosen key is derived which concludes that this algorithm provides security against Brute-force attack and the selected identifiable key can play role of secret key against Brute-force attack.

The paper is organized as follows. In Section II, a brief description of the approaches involved in the analysis of the proposed algorithm and section III, discusses the overview of Non-Linear Shift Register and Logistic map used for developing the algorithm. Section IV, an algorithm of the proposed encryption method is presented. In Section V, simulated analysis results are cited in tabulated form. Then in Section VI, the conclusion derived from simulated analysis is discussed and it is shown that this method provides security against Brute-force attack.

## 2. CRYPTANALYTIC PROCEDURES

**A. Output Equality:** The output equality describes that - For the same inputs and initial condition, transmitter system is parameterized at different values of parameter taken from the existing domain of parameter space, if the output response of the system obtained after some value of iteration, parameterized at a particular value coincides with the output response of the same system parameterized at some other value of parameter within the domain for the same number of iteration, then both the parameters are said to be equal and identifiable. The system is said to possess unique solution at that particular value of parameter and the system is said to be structurally identifiable.

There exists a connection between uniqueness in the secret parameters (acting as key) and identifiability concept, which, reduces the probability of finding actual parameter by the eavesdropper. If parameter of the transmitter is identifiable, it is more difficult for the eavesdropper to find it by a brute force attack. Consequently, this parameter may be a good candidate to play the role of the secret key against a brute force attack. If the parameter is not identifiable, the eavesdropper has a higher favorable chance to find it by a brute force attack. Thus, this parameter vector is a bad candidate to play the role of the secret key against brute force attack.

**B. Plaintext sensitivity Test:** It is the percentage of change in bits of cipher text obtained after encryption of plaintext, which is derived by changing single bit from the original plaintext from the bits of cipher text obtained after encryption of original plaintext. With the change in single bit of plaintext, there, must be ideally 50% change in bits of cipher text to resist differential cryptanalysis (chosen-plaintext attack) and statistical analysis, corresponds to plaintext sensitivity test.

**C. Key sensitivity Test:** Key sensitivity is the percentage of change in bits of cipher text obtained after encryption of plaintext using key, which is flipped by single bit from the original key, from bits of cipher text obtained after encryption of plaintext using original key, which requires ideally 50% change in cipher text bits to resist linear and statistical attacks.

**D. Known plaintext attack:** For implementing this attack on developed cryptosystem it is assumed that the opponent knows everything about the algorithm, he/she has the cipher text and some portion of plaintext. With this much information, the opponent tries to find out the secret key.

**E. Key space analysis:** The key space is an important part of analysis that corresponds to the range of region in which the behavior of chaotic system used in cipher is chaotic in nature.

To resist common attacks, the designed cryptosystem should have the confusion property. To achieve the confusion property, statistical properties of the cipher text, such as distribution, correlation and differential probability of the cipher text should be independent of the exact value of the key and of the plaintext. With the increase of randomness of cipher text, confusion strengthens. No pattern should involve in the cipher text for secured cryptosystem. Confusion is intended to make the relationship between cipher text and plaintext statistically independent.

Plaintext sensitivity and key sensitivity describes the diffusion property of the system. Diffusion refers spreading out of the influence of single plaintext digit over many cipher text digits so as to hide the statistical structure of the plaintext. These methods together are also known as Avalanche effect.

## 3. Functions Used For Designing Cryptosystem

A. Non-linear Feedback Register: NLFSR (Non-Linear Feedback Shift-register) is a common component in modern stream ciphers, especially in RFID and smartcard applications. NLFSRs are known to be more resistant to cryptanalytic attacks than Linear Feedback Shift Registers (LFSR's), although construction of large NLFSRs with guaranteed long periods remains an open problem. A NLFSR is a shift register whose current state is a non-linear function of its previous state. The NLFSR used in this paper is shown in fig 2.

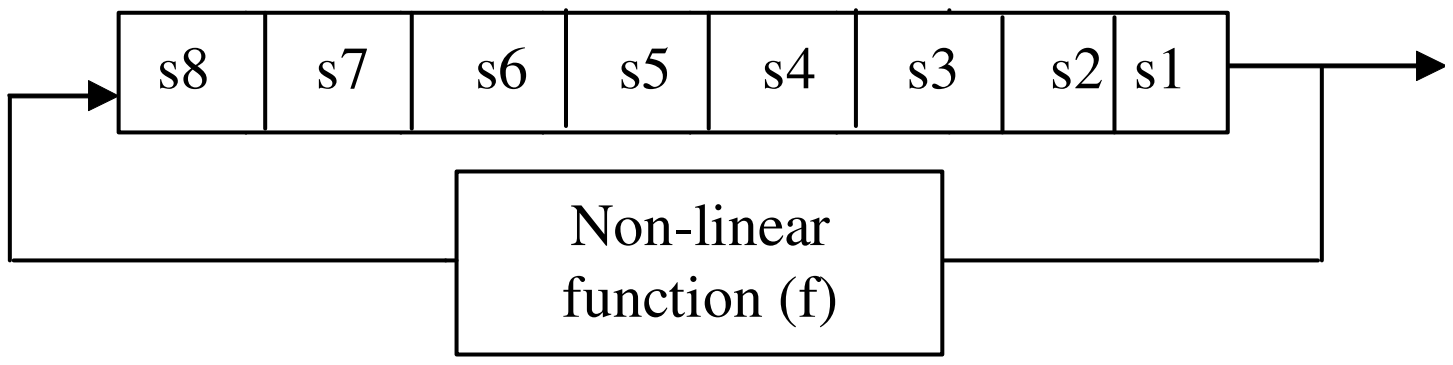

Figure 2. NLFSR using 8-bit shift register

**B. Logistic map:** The logistic map is a polynomial mapping of degree 2, it takes a point, in a plane and maps it to a new point using following expressions:

$$x(k+1) = rx(k)\left[1 - x(k)\right]$$

Where, map depends on the parameter r. From r = 3.57 to r = 4, the map exhibits chaotic behavior which is shown in fig 3(a) and (b).

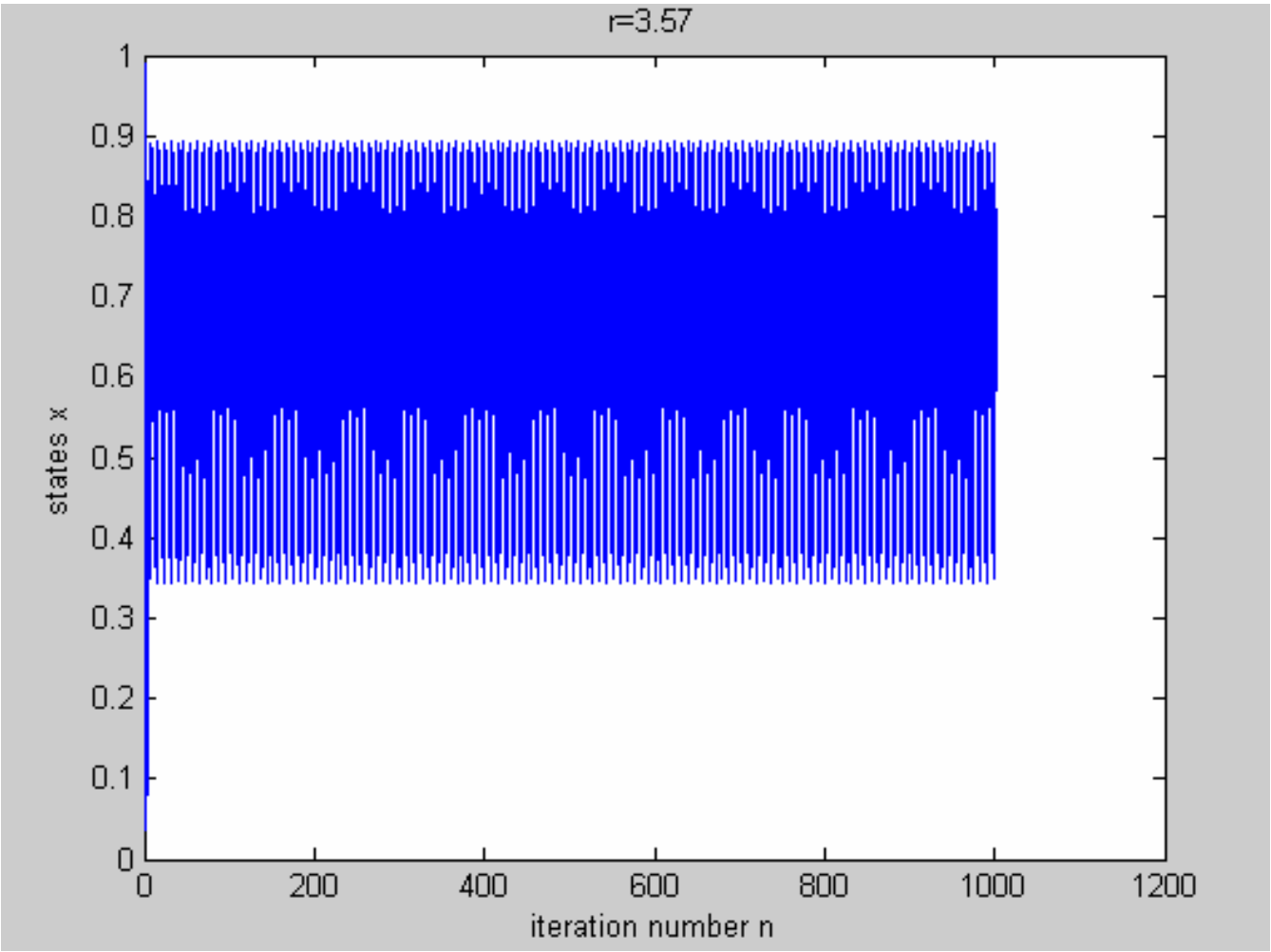

Figure 3. (a) Plot of Logistic map for r = 3.57, x (0) = 0.99, n = 1000.

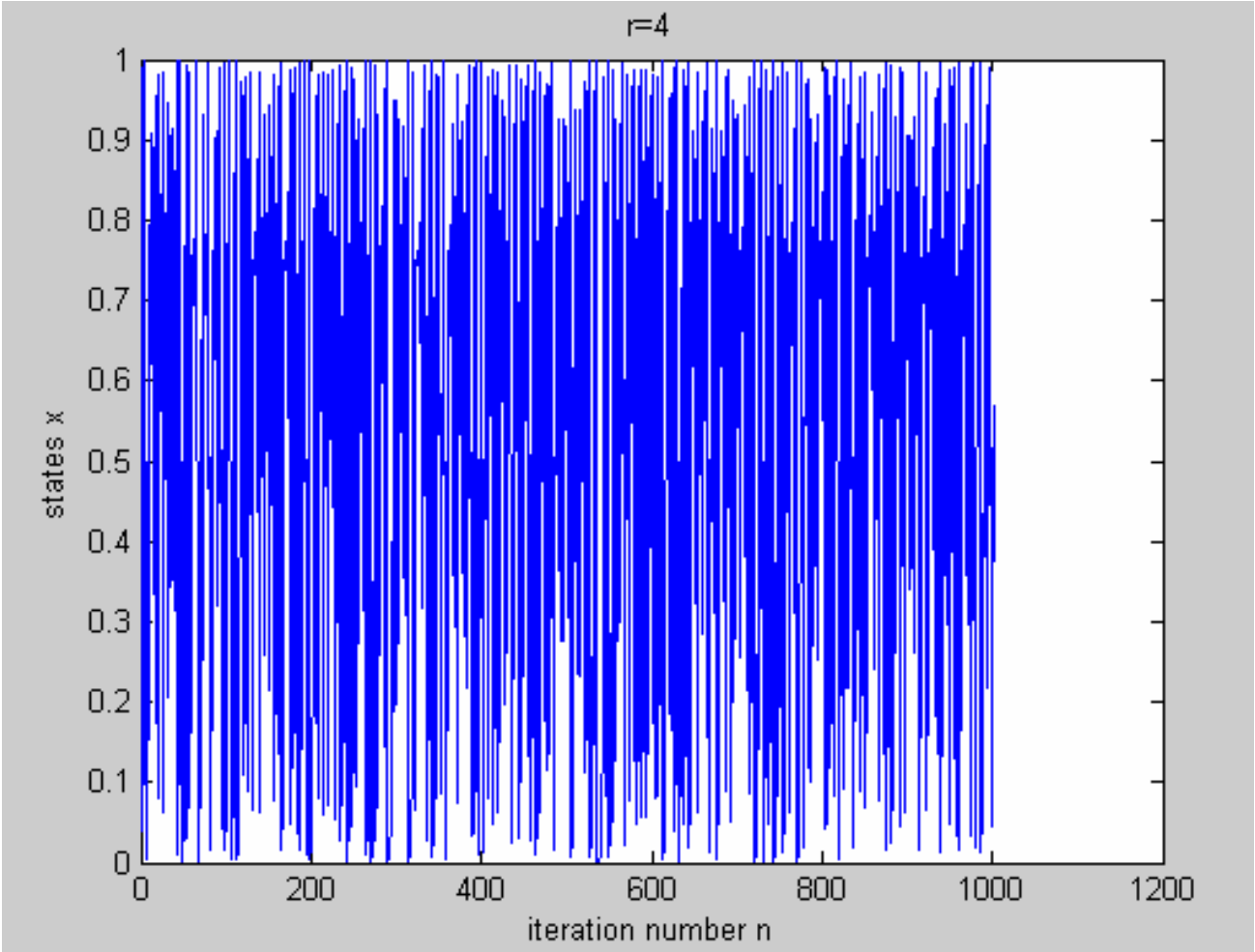

Figure 3. (b) Plot of Logistic map for r = 4.0, x (0) = 0.99, n = 1000.

## 4. PROPOSED ENCRYPTION METHOD

Algorithm for the proposed encryption procedure is discussed in brief way.
Encryption Algorithm

Step 1. Read plaintext, key parameter, initial conditions, and length of text, index of length and index for iteration.

Step 2. Read, matrix K.

Step 3. Read value of x = 128.

Step 4. Convert plaintext into ASCII form.

Step 5. Calculate mod (K p, x).

Step 6. For the length of plaintext, do following steps:

(a) (i) Read input number, which must be positive.
(ii) Convert it into binary form.
(iii) Read length of binary number.
(iv) Read register
(v) Load register with the input number.
(vi) Read non-linear function for feedback to first bit of register
(vii) Read index of length of number.
(viii) For the length of number, content of shift register is shifted
(viii) Loop continues till the length becomes unity.
(ix) Go to (a.)
(x) Convert output binary number into decimal number.
(xi) Read output number and store it.
(b) Iterate the logistic map for any number of iterations and obtain the state.
(c) Calculate, intermediate encrypted plaintext.
(d) Again iterate logistic map for any number of iterations and obtain the state.
(e) Mix the state with non-linear function and obtain the ciphertext.

Step 7. Read output text.

Step 8. Output is plotted.

Step 9. Convert output into characters.

Step 10. Read cipher text.

Decryption algorithm is just reverse of encryption.

## 5. ANALYSIS RESULT

Cryptanalysis plays most important role in the development and advancement of cryptography. In this section twenty different texts have been chosen as plaintext to the developed method and are analyzed for plaintext sensitivity, key sensitivity, Identifiability and known plaintext attack for different keys selected from domain of key space. The analysis results are presented in tabular form obtained from simulation of the cryptanalytic procedures. As we know that logistic map is chaotic for the value of parameter ‘r’= 3.57 to 4.0, hence key space of the method is also limited to chaotic region, which is very less compared to $2^{100}$. The chaotic system is sensitive to a small change in initial condition or parameters, which shows that if value of key is incremented up to the precision of $10^{-9}$, then also the system shows a drastic change in its output behavior, which enlarges the key space still not sufficient to resist brute-force attack because to resist this attack it is required to have precision of $10^{-15}$. The key space for this algorithm is determined approx. to be $47 \times 10^{7}$.

Table 1. Analysis result

NI – Non-Identifiable; I – Identifiable; R – Robust; p [p1 p2… p n] – First ‘n’ characters of available plaintext string; NR – Not robust;

| S.No. | Plaintext | Key value | Cipher text | Plaintext sensitivity (in %) | Key sensitivity (in %) | Domain for key With increment =0.0001 | Identifiability of key for iteration value =1 or 2 | Robustness against known plaintext attack. | Whether key can act as secret key against Brute Force attack? |
|---|---|---|---|---|---|---|---|---|---|
| 1. | What is your name? | 3.6424 | -§Án_ª¶¼X<Ó¬sÊmÄ | 9.2105 | 16.4474 | (3.57,3.77) | I | R for p=[p1 p2 ...p5] | YES |
| 2. | I am going to market. | 3.7328 | \|\| vfª(ò´(¬ô6ýR2& | 10.7955 | 17.0455 | (3.57,3.77) | I | R for p=[p1 p2 ...p19] | YES |
| 3. | My college name is s.s.c.e.t. | 3.7455 | ,Üü,¢&-ñØ&Z®Ê\Ö:Ö:þÚp2& | 5.4167 | 13.3333 | (3.57,3.77) | I | R for p=[p1 p2 ... p27] | YES |
| 4. | Hello!how are you? | 3.7694 | ¶¢>RÌ~Ú^Ê$ôjÄ | 7.8947 | 9.8684 | (3.57,3.77) | I | R for p=[p1 p2 ...p15] | YES |
| 5. | Sita is singing very well. | 3.8544 | BÛ¹⁄₄-eÂ=ßRf1àdØùFì;d*p | 6.4815 | 12.9630 | (3.66,3.86) | I | R for p=[p1 p2 ...p19] | YES |
| 6. | Ram scored 98 marks in Maths. | 3.8551 | L² ½jdSÜ*• ¢æ%<Äc&;$ÍèÅ<-±& | 5.8333 | 14.5833 | (3.76,3.96) | I | R for p=[p1 p2 ...p19] | YES |
| 7. | Jaycee publication. | 3.8529 | =txý®&Ó`Fàâ$# | 14.3750 | 23.1250 | (3.77,3.97) | I | R for p=[p1 p2 ... p15] | YES |
| 8. | Thank you,sir. | 3.8641 | ñÖæN;EP¼0eé:o | 10.8333 | 20 | (3.78,3.98) | I | Rfor p=[p1 p2 ...p11] | YES |
| 9. | The match was very exciting. | 3.9065 | ÝïÌWýÛ*¡)=ÈEÓPá?I×4• löK | 6.4655 | 22.8448 | (3.785,3.985) | I | R for p=[p1 p2 ... p19] | YES |
| 10. | I will be leaving at 9p.m. | 3.9189 | O– ¯Ø,§8 -üÊ¨0 Lfo | 6.4815 | 12.9630 | (3.79,3.99) | I | R for p=[p1 p2 ... p19] | YES |
| 11. | How are you? | 3.5700 | §À ä_Ï"à»ø´ | 12.5 | 11.4 | (3.57,3.77) | I | R for p = [p1p2...p12 ] | YES |
| 12. | Meet me after 5p.m. | 3.5701 | &36 ëþ ‚=f ÝZØLep | 9.8684 | 9.2105 | (3.5701,3.7701) | I | R for p = [p1p2...p20] | YES |
| 13. | I have a gift for you. | 3.5709 | Ä ÒPý Î" üìÎë "•!Ò% | 8.5227 | 6.8182 | (3.5709,3.7709) | I | R for p = [p1p2...p17] | YES |
| 14. | We will go for walk. | 3.5800 | Ä ÒPý Î" ükÎë "•!Ò% | 8.5227 | 10.2273 | (3.5709,3.7700) | I | R for p = [p1p2...p21] | YES |
| 15. | Study different papers. | 3.5801 | =þ2æ9 ð nÀt ÞüF- F rYp | 8.6957 | 8.1522 | (3.5700,3.7600) | I | R for p = [p1p2...p23] | YES |
| 16. | How to do analysis? | 3.5809 | §À I¢ è]¾&`YB èYÄ | 7.8947 | 14.4737 | (3.5700,3.7600) | I | R for p= [p1p2...p19] | YES |
| 17. | Hai! Where are you going? | 3.5900 | Økrk8 $ÞNÑ>ÿ^É$ó*cè;f/Èv | 6.5 | 12.00 | (3.5700,3.7600) | I | R for p = [ p1p2...p21] | YES |
| 18. | Dolly, are you coming with me? | 3.5901 | ®i¸¹ü,é^É$ô YÞ%r >Õx \|*äÅÒJÚA | 6.25 | 12.0833 | (3.5700,3.7600) | I | R for p = [p1p2...p25] | YES |
| 19. | Children are playing in park. | 3.5905 | ı Â1 Ý0ø<üLÖðáÚZbo\|#Î ɛ &ôL | 5.8333 | 9.5833 | (3.5700,3.7600) | I | R for p= [ p1p2... p25] | YES |
| 20. | I shall go to cinema | 3.600 | ò)°7• ± 6 'dV Ö %L | 8.3333 | 13.6905 | (3.5700,3.7600) | I | NR | YES |

It is concluded from the results cited in the given table 1 that plaintext sensitivity of the method ranges from 5% to 15% and key sensitivity is ranging from 6 % to 23%, which concludes that the method may or may not be weak against linear, statistical and differential attacks. For almost all the selected keys, method resist known plaintext attack for available first five characters of plaintext string except the last one. If available character string of plaintext, which may be in any number are not the starting characters of plaintext then in such situation method proves to resist the attack for all the selected keys. Different texts analyzed for known-plaintext attack for different keys requires different numbers of available characters of plaintext to find out the secret key as mentioned in the table ,which concludes that strength of the algorithm against the attack depends on the length of text and strength of key. Each of the key gives different performance, thus enhance the unpredictability and accordingly it can be used for particular applications. Conclusion about the identifiability of the all chosen key is derived for the given specifications in the above table which concludes that method can provide security against Brute-force attack and the selected identifiable key can play role of secret key against Brute-force attack. It is seen that length of cipher text is equal to the length of plaintext. Time taken is less and tolerable. The space between the characters in plaintext is encrypted. In encrypted text, it is seen that the letter of plaintext, that repeats is not encrypted as same encrypted text, which avoids any pattern to exist in cipher text, thus shows randomness of cipher text.

A complete file as shown in fig 4 is encrypted using the procedure as shown in fig 5, which concludes that the method is capable of encrypting and decrypting a complete file successfully.

```
C:\Documents and Settings\ussr\My Documents\MATLAB\plaintext.txt*
Branch and subject code-CSVTU
Course Name: ME (PT), ME - M Tech
Sr. No. Branch Code Branch name Subject Code    Subject
MACHINE DESIGN
1   Design  48  548111 (37) Tribology
2   Design  48  548112 (37) Mechanical Vibration
3   Design  48  548113 (37) Advanced Dynamics of Machine
4   Design  48  548114 (37) Theory of Elasticity & Plasticity
5   Design  48  548131 (37) Optimization Techniques (Elective I)
6   Design  48  548211 (37) Finite Element Methods
7   Design  48  548212 (37) CAD/CAM Application
8   Design  48  548213 (37) Advanced Machine Tool Design
9   Design  48  548214 (37) Advanced Mechanism
10  Design  48  548231 (37) Experimental Stress Analysis (Elective II)
11  Design  48  548311 (37) Robotics
12  Design  48  548312 (37) Fatigue & Creep
PRODUCTION ENGINEERING
1   Production Engg.    42  542111 (37) Advanced Manufacturing Engineering
2   Production Engg.    42  542112 (37) CAD/CAM Applications
3   Production Engg.    42  542113 (37) Production & Materials Management
4   Production Engg.    42  542114 (37) Maintenance Engineering
5   Production Engg.    42  542131 (37) Applied Fuzzy Logic & Fuzzy Sets
6   Production Engg.    42  542211 (37) Machine Tools Engg.
7   Production Engg.    42  542212 (37) Robotics
8   Production Engg.    42  542213 (37) Quality Control & Reliability
9   Production Engg.    42  542214 (37) Measurement System Analysis
10  Production Engg.    42  542232 (37) Productivity Management
```

Figure 4. Plaintext file

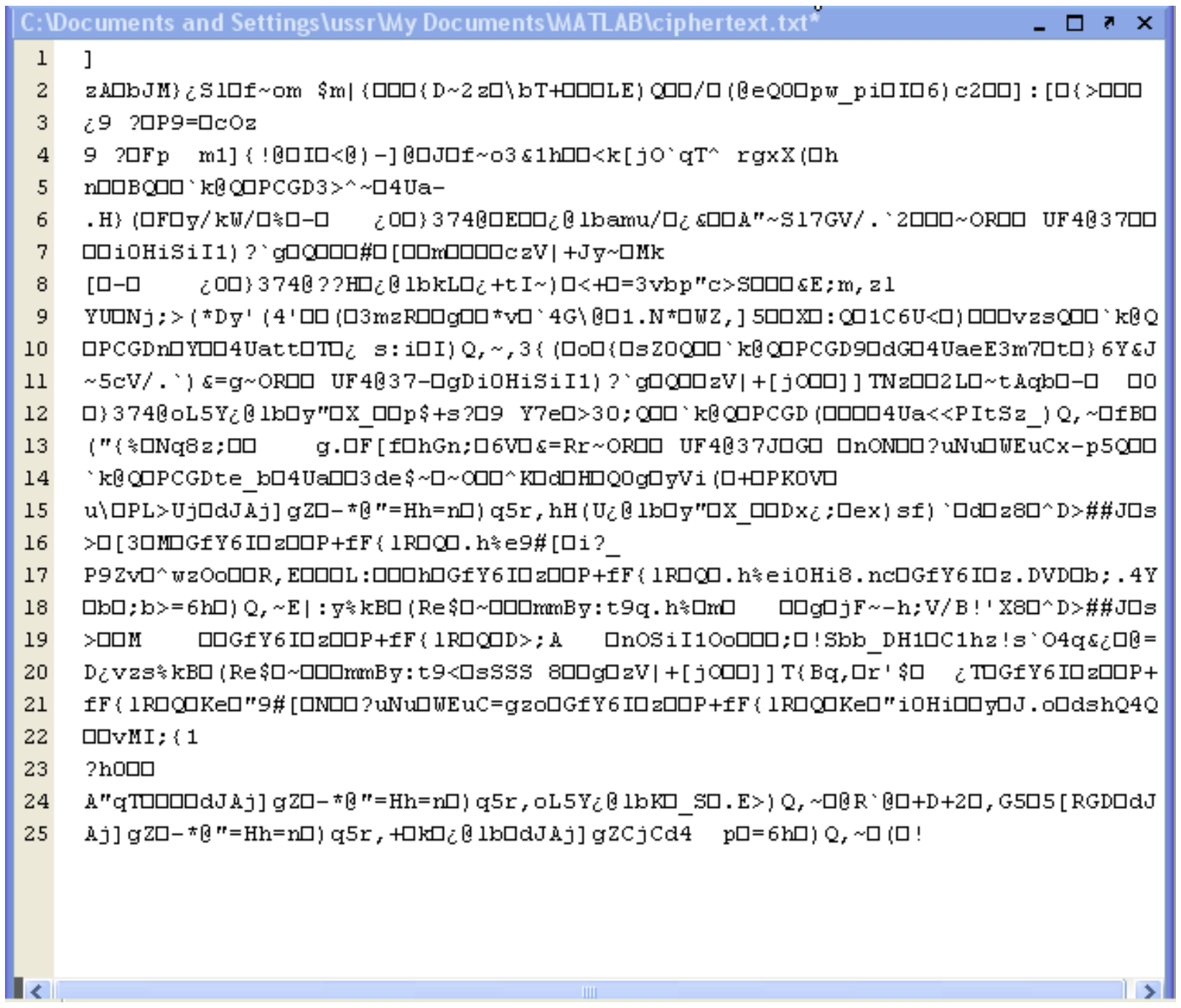

Figure 5. Encrypted file

## 6. CONCLUSION

The paper proposes a hybrid message embedded scheme which consists of hill cipher method combined with message embedded scheme. Message-embedded scheme using non-linear feedback shift register as non-linear function and 1-D logistic map as chaotic map is modified and analyzed and tested for avalanche property and strength against known plaintext attack and brute-force attack. Parameter of logistic map acts as a secret key. We know that the minimum key space to resist brute-force attack is 2100 and it is seen that key space of the discussed method is lesser than 2100. But the identifiability test concludes that the scheme consists of identifiable keys which are sufficient condition to resist brute-force attack for chaotic ciphers.

A complete file can be encrypted and decrypted successfully by the method that can provide security against brute force attack. It is shown that the scheme has an average key sensitivity and most of the analyzed keys have strength against known plaintext attack.

## ACKNOWLEDGEMENTS

The authors would like to thank the anonymous reviewers for their valuable suggestions and the proposed references.

## Authors

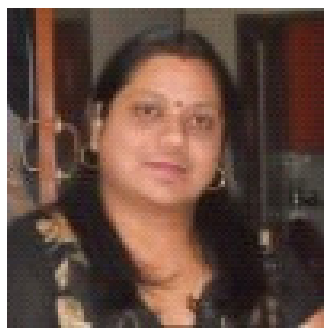

Mina Mishra, is pursuing Ph.D. (Engg) from Nagpur University, Maharashtra, India. She received M.E. degree specialization in communication in the year 2010. Her research area covers chaotic systems, chaotic cryptology, network security and secure communication.

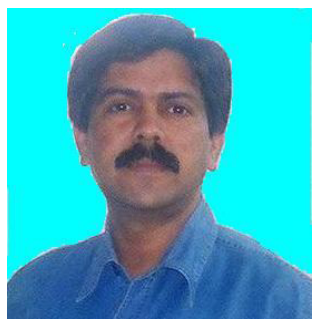

Vijay H. Mankar received M. Tech. degree in Electronics Engineering from VNIT, Nagpur University, India in 1995 and Ph.D. (Engg) from Jadavpur University, Kolkata, India in 2009 respectively. He has more than 16 years of teaching experience and presently working as a Lecturer (Selection Grade) in Government Polytechnic, Nagpur (MS), India. He has published more than 30 research papers in international conference and journals. His field of interest includes digital image processing, data hiding and watermarking.